\documentstyle[amsfonts]{europhys}
\def\stars{\bigskip\centerline{***}\medskip}

\newif\ifboo \boofalse

\def\Review#1{\boofalse{\it #1},}
\def\Name#1{{\sc #1},}
\def\Vol#1{\ifboo Vol. {\bf #1}\else{\bf #1}\fi}
\def\Year#1{\ifboo #1\else(#1)\fi}

\def\Page#1{\ifboo {\rm p. #1}\else{\rm #1}\fi}
\newcommand{\cft}{conformal field theory }
\newcommand{\cor}{\overline{\langle \sigma(0)\sigma(R)\rangle^p}}
\newcommand{\opx} {{\cal O}_p (x)}
\newcommand{\nn}{\nonumber}
\begin{document}
%
%%%   The headers.
%
%%%   These three macros are to have correct headings in your paper.
%%%   You shall omit all the arguments in the two macros `\euro{}{}{}{}'
%%%   `\Date{}' and fill in `\shorttitle{}'. 
%%%   If there is more than one author in the 
%%%   \shorttitle macro, use the macro \etal after first author's name
%%%   to obtain the correct heading.
%
\euro{}{}{}{}
\Date{}
\shorttitle{M.-A LEWIS Spin-spin moments for the ferromagnetic random bond Potts model}
%
%%%  The title, the Author(s) and the affiliation(s)
%
%%%   The title is set in bold (initial word only is capitalized).
%%%   Mathematical expressions and formulas within the title shall be left
%%%   in light face. Initial(s) of the first name(s) are followed by the
%%%   author(s)'s last name(s). If the authors have different affiliations,
%%%   the name must be followed by one or more \inst{number} each referring
%%%   to one of the addresses to appear in the following macro \institute.
%%%   Other items like `Present address' or `email' may be added by putting
%%%   a `\footnote' after the last \inst{number}.
%%%   Begin each address with \inst{number}; the end of an address is \\;
%%%   \\ can also be used to break a line.
%
\title{Higher moments of spin-spin correlation functions for the \\ ferromagnetic random bond Potts model}
\author{Marc-Andr\'e Lewis\footnote{E-mail:lewism@lpthe.jussieu.fr}}
\institute{Laboratoire de Physique Th\'eorique et des Hautes Energies,\footnote{ 
 Unit\'e associ\'ee au CNRS URA 280} \\
         Universit\'es Pierre et Marie Curie (Paris VI) et Denis Diderot (Paris VII),\\
Boite 126, Tour 16, 1er \'etage,\\         
4 pl. Jussieu, 75251 Paris CEDEX 05, FRANCE}
     
%
%%%    The `\maketitle' macro needs the following macro:    \rec{}{}
%%%    to be left empty.
%
\rec{}{}
%
%%%   Physics Abstracts Classification.
%
%%%   There are two macros: the first one `\pacs{}' makes the PACS 
%%%   environment,the second one `\Pacs{}{}{}' can be used for each
%%%   classification you need.
%%%   To create the subject index of the volume it is important to divide
%%%   the classification numbers into the three different arguments like
%%%   in the following examples 
%
\pacs{
\Pacs{64}{60Ak}{Renormalization group studies of phase transitions}
\Pacs{11}{25Mj}{Conformal field theory; algebraic structures}
      }
\maketitle
%
%%%   ! Don't forget this command to format the title page of your article!
%
%%%   The Abstract
%
\begin{abstract} Using \cft techniques, we compute the disorder-averaged
 $p$th power of the spin-spin correlation function ($\cor,\,p\in\mathbb Z$) 
for the ferromagnetic random bond Potts model.  We thus generalize the calculations of Dotsenko, Dotsenko and Picco, where the case $p=2$ was considered, and of Ludwig, where first-order computations where made for general $p$.  Perturbative calculations  are  made up to the second order in $\epsilon$ ($\epsilon$ being proportional to the central charge deviation of the pure model from the Ising model value).  The explicit dependence of the correlation function on $p$ gives an upper bound for the validity of the $\epsilon$-expansion, which seems to be valid, in the three-states case, only if $p\le 4$.
\end{abstract}
%
%
%%%   Main text
%
%--------------------------------------------------------------------
Since the first calculations made by Ludwig \cite{Lu}, a lot of attention was
given to the study of the random bond Potts model.  It was established that
the introduction of randomness changes the critical behaviour of the system, as predicted by the Harris criterion.  Using perturbative \cft techniques \cite{BPZ} and $\epsilon$-regularization which consists here in a shift of central charge from the Ising model value,  first order \cite{Lu} and second order calculations \cite{DPP} clearly established the existence of fixed points in the renormalization group flow.  In fact, there exists two different fixed point solutions: one with replica symetry (RS) and another where this symetry is broken (RSB).  Recent results by  Dotsenko, Dotsenko and Picco \cite{DDP} support the RS fixed point critical behaviour of the random bonds Potts model.  To compare both schemes, they compute the disorder-averaged second moment of the spin-spin correlation function $\overline{\langle \sigma(0)\sigma(R)\rangle^2}$ with broken and unbroken replica symetry.  Numerical simulations for the 3-states and 4-states models don't shown significant deviation from the replica symetric solution.

However, the observed higher moments of correlation functions seem to be in contradiction with the values predicted by the RS $\epsilon$-expansion calculations \cite{Ta}. In this letter, we will compute the disorder-averaged $p$-th power of the spin-spin correlation function ($\cor$) in the replica symetric case.  The explicit dependence of this quantity on $p$ shows how the expansion validity breaks down for sufficiently large $p$.  We find, for the 3-state Potts model, that the expansion is valid only if $p\le 4$, thus confirming the difference between observed and predicted values for high moments.

The partition function of the nearly-critical $q$-states random bond Potts model, is well known to be of the form
\begin{equation}
\label{Z}
Z(\beta) =  \mbox{Tr } \exp\{-H_0-H_1\},
\end{equation}
where $H_0$ is the Hamiltonian of the \cft corresponding to the $q$-states Potts model with coupling constant $J_0$ the same for each bond.  The Hamiltonian $H_1$, being the deviation from the critical point induced by disorder is of the form
\begin{equation}
H_1= \int d^2x \,\tau(x)\epsilon(x),
\end{equation}
where $\tau(x)\sim\beta J(x) - \beta_c J_0$ is the random temperature parameter. The theory is defined on the whole plane.  We shall assume, for simplicity, that $\tau(x)$ has a gaussian distribution for each $x$, with
\begin{eqnarray}
\overline{\tau(x)} &=& \tau_0 \,= \frac{\beta-\beta_c}{\beta_c}\\
\overline{(\tau(x)-\tau_0)(\tau(x')-\tau_0)} &=& g_0\, \delta^{(2)}(x-x')
\end{eqnarray}

The usual way of averaging over disorder is to introduce replicas, that is, $n$ identical copies of the same model for which:
\begin{equation}
(Z(\beta))^n=\mbox{Tr }\exp\{-\sum_{a=1}^n H_0^{(a)} - \int d^2x \,\tau(x)\sum_{a=1}^n \varepsilon_a (x)\}.
\end{equation}
Taking the average over disorder, by performing gaussian integration, one gets
\begin{equation}
\overline{(Z(\beta))^{n}}=\mbox{Tr }\exp\{-\sum_{a=1}^{n}H_{0}^{(a)}-
\tau_{0}\int d^{2}x\sum_{a=1}^{n}\varepsilon_{a}(x)+g_{0}\int 
d^{2}x\sum_{a\neq b}^{n}\varepsilon_{a}(x)\varepsilon_{b}(x)\}.
\end{equation}
This is a field theory of $n$ coupled models with coupling action given by
\begin{equation}
H_{\mbox{int}}=-g_{0}\int d^{2}x\sum_{a\neq
b}^{n}\varepsilon_{a}(x)\varepsilon_{b}(x). 
\end{equation}
Note that only non-diagonal terms are kept since diagonal ones can be included in the Hamiltonian $H_0$.  Moreover, they can be shown to have irrelevant contributions, since their OPE consist of the identity plus terms that are irrelevant at the pure fixed point. We now turn our attention to the $p$-th moment of the spin-spin correlation function $\cor$.  In terms of replicas, it can be written as
\begin{eqnarray}
\cor &=& \lim_{n\rightarrow 0} \frac{(n-p)!}{n!} \sum_{a_1\ne a_2 \cdots \ne a_p}^n \langle \sigma_{a_1}(0)\sigma_{a_1}(R) \cdots \sigma_{a_p}(0)\sigma_{a_p}(R)\rangle\nn\\
     &=&  \lim_{n\rightarrow 0} \frac{(n-p)!}{n!\,p!}  \langle \sum_{a_1\ne\cdots\ne a_p}^n \sigma_{a_1}(0)\cdots \sigma_{a_p}(0) 
\sum_{b_1\ne \cdots \ne b_p}^n \sigma_{b_1}(R)\cdots\sigma_{b_p}(R)\rangle.
\end{eqnarray}
The operator to be renormalized is then
\begin{equation}
{\cal O}_p (x) \equiv \sigma_{a_1}(x)\sigma_{a_2}(x)\cdots\sigma_{a_p}(x),\qquad a_1\ne a_2 \cdots \ne a_p,\,\, 1\le a_i \le n,
\end{equation}
perturbed by the interaction term;
\begin{equation}
\tilde\opx \equiv {\cal O}_p \exp\{-H_{\mbox{int}}\} = {\cal O}_p \left( 1-H_{\mbox{int}}+\frac{1}{2}(H_{\mbox{int}})^2-\cdots\right).
\end{equation}
We will define the amplitude $Z$, for which we will derive RG equations, as
\begin{equation}
\label{renorm}
\tilde\opx = Z {\cal O}_p(x).
\end{equation}
The task at hand is thus to rewrite $\tilde\opx$ in the form (\ref{renorm}) by doing all possible contractions and operator algebra.  We will compute $Z$ up to the second order in $g_0$.  To do so, we use the Coulomb gas formulation of minimal conformal field
theories \cite{DF1}.  In this formalism, the central charge of the theory is written as
\begin{eqnarray}
c&=&1-24\alpha_0^2 \\
\alpha_{\pm} &=& \alpha_0 \pm \sqrt{\alpha_0^2+1} \qquad \alpha_+\alpha_- = -1.
\end{eqnarray}
For the Ising model, $\alpha_+^2=4/3$ and $c=1/2$, while for the 3-states Potts model, $\alpha_+^2=\frac{6}{5}$ and $c=\frac{4}{5}$.  For a generic model, we will write $\alpha_+^2=\frac{4}{3}-\epsilon$, so that $c=\frac{1}{2}+\frac{21}{8}\epsilon+{\cal O}(\epsilon^2)$.  In particular, $\epsilon=\frac{2}{15}$ corresponds to the 3-states Potts model.  For Potts models, the energy operator $\varepsilon(x)$ is the primary field $\Phi_{1,2}$ so that its conformal dimension is
\begin{equation}
\Delta_{\varepsilon} = \Delta_{1,2} + \bar{\Delta}_{1,2} = \frac{(\alpha_-+2\alpha_+)^2 - (\alpha_-+\alpha_+)^2}{2} = 1-\frac{3}{2}\epsilon.
\end{equation}

We shall often use the spin and energy operators product expansion
\begin{equation}
\sigma(x)\varepsilon(y) = \frac{D}{|x-y|^{\Delta_\varepsilon}}\sigma(x)+ \mbox{finite contributions},
\end{equation}
where $D$, the operator algebra coefficient, is known to be $\frac{1}{2}+{\cal O}(\epsilon^2)$ \cite{DF3}.  One can get rid of the finite terms by projecting correlations functions on $\sigma(\infty)$.

Renormalization group equations will be derived by integrating from a cut-off of 1 (in lattice spacing units) to a new one $a$ ($a\gg 1$).  First order calculations are straightforward.  Since operators with different replica indexes have zero product expansion, there are $p(p-1)$ possible contractions, that is
\begin{eqnarray}
-\opx H_{\mbox{int}} &=& \sigma_{a_1}(x)\cdots\sigma_{a_p}(x) g\int d^{2}y\sum_{c\ne d}^n \varepsilon_{c}(y)\varepsilon_{d}(y)\nn\\
&\rightarrow& \sigma_{a_1}(x)\cdots\sigma_{a_p}(x)g\,p(p-1)\int_{1<|y-x|<a} d^2y \langle \sigma(x)\varepsilon(y)\sigma(\infty)\rangle^2\nn \\
&=& \sigma_{a_1}(x)\cdots\sigma_{a_p}(x)g\,p(p-1)\int_{1<|y-x|<a} d^2y \frac{D^2}{|x-y|^{2\Delta_\varepsilon}}\nn \\
&=& \opx p(p-1)\,g\,\frac{2\pi D^2}{3\epsilon} a^{3\epsilon}. 
\end{eqnarray}
So, the first order corrections to $Z$ are
\begin{equation}
\delta Z^{(1)} = Z p(p-1)\,g\,\frac{2\pi D^2}{3\epsilon} a^{3\epsilon}.
\end{equation}
Second order calculations require more work.  There are five different types of contractions possible; three of them occur if $p\ge 2$, the fourth if $p\ge 3$ and finally the fifth if $p\ge 4$.  The first three diagrams were computed in \cite{DDP}, and, for generic $p$, only combinatorial factors are modified.  We will only give their expression and concentrate on the computation of the two last diagrams.  The first diagrams give the contributions
\begin{eqnarray}
D_1^{(2)} &=& \opx p(p-1)(n-2) g^2\frac{4\pi^2 D^2}{9\epsilon^2}(1+\epsilon K) a^{6\epsilon}\\
D_2^{(2)} &=& \mbox{finite contributions}\\
D_3^{(2)} &=& \opx p(p-1)g^2 \left(\frac{4\pi^2 D^4}{9\epsilon^2} - \frac{\pi^2}{36\epsilon}\right)a^{6\epsilon},
\end{eqnarray} 
where $K=6\log 2$. We only consider the divergent part of the diagrams since these are the only ones appearing in the RG equations.  The fourth diagram expression is given by
\begin{eqnarray}
D_4^{(2)} &=& \opx \frac{p!}{(p-3)!} g^2 \int\int d^2 y \,d^2 y'\, \langle \sigma(0)\varepsilon(y)\varepsilon(y')\sigma(\infty)\rangle \langle \sigma(0)\varepsilon(y)\sigma(\infty)\rangle  \langle \sigma(0)\varepsilon(y')\sigma(\infty)\rangle \nn \\
         &=&\opx \frac{p!}{(p-3)!} g^2  D^2 \int\int d^2 \,y d^2 y'\, |y|^{-\Delta_{\varepsilon}}|y'|^{-\Delta_{\varepsilon}}\langle \sigma(0)\varepsilon(y)\varepsilon(y')\sigma(\infty)\rangle.
\end{eqnarray}
A trivial change of variable and the use of the fact that 
$$\langle \sigma(0)\varepsilon(y)\varepsilon(y')\sigma(\infty)\rangle = \lambda^{2\Delta_{\varepsilon}} 
\langle \sigma(0)\varepsilon(\lambda y)\varepsilon(\lambda y')\sigma(\infty)\rangle$$
leads to
\begin{eqnarray}
D_4^{(2)} &=& 2\opx p(p-1)(p-2) g^2 \int d^2 y' |y'|^{2-4\Delta_{\varepsilon}} \int d^2 y |y|^{-\Delta_{\varepsilon}} \langle \sigma(0)\varepsilon(1)\varepsilon(y)\sigma(\infty)\rangle \nn \\
&=& 2\pi D^2 \frac{a^{6\epsilon}}{6\epsilon} \int d^2 y |y|^{-\Delta_{\varepsilon}} \langle \sigma(0)\varepsilon(1)\varepsilon(y)\sigma(\infty)\rangle.
\end{eqnarray}
The calculation of this integral is done with the use of the techniques described in \cite{DDP}.  One gets
\begin{eqnarray}
 D_4^{(2)} &=& \opx p(p-1)(p-2) g^2 \frac{\pi^2 D^2}{18\epsilon^2} (8+\epsilon\alpha)a^{6\epsilon},
\end{eqnarray}
with $\alpha = 33 - \frac{29\sqrt{3}\pi}{3}$.

The calculation of $D_5^{(2)}$ is simpler.  The diagram consists of four $\sigma\epsilon$ contractions:
\begin{eqnarray}
D_5^{(2)} &=& \frac{1}{2} \opx p(p-1)(p-2)(p-3) g^2 \left(\int d^2 y \frac{D^2}{|y|^{2\Delta_{\varepsilon}}}\right)^2 \nn\\
&=& \opx p(p-1)(p-2)(p-3) g^2 \frac{2\pi^2 D^4}{9\epsilon^2} a^{6\epsilon}
\end{eqnarray}
 
Collecting all results, we get the second order correction to $Z$:
\begin{eqnarray}
\delta Z^{(2)} &=& Z g^2 p(p-1)a^{6\epsilon} \left( (n-2)\frac{4\pi^2 D^2}{9\epsilon^2}(1+\epsilon K)  + 
\left(\frac{4\pi^2 D^4}{9\epsilon^2} - \frac{\pi^2}{36\epsilon}\right)\right. \nn \\
 &&\,\,\left.+ (p-2)\left(\frac{\pi^2 D^2}{18\epsilon^2} (8+\epsilon\alpha) + (p-3)
\frac{2\pi^2 D^4}{9\epsilon^2}\right)\right)
\end{eqnarray}
We can now write the RG equation for $Z$ ($\xi\equiv \log a$):
\begin{equation}
\frac{dZ}{d\xi} = a \frac{dZ}{da} = Z \left( A(p,\epsilon) g(a) a^{3\epsilon} + B(p,\epsilon) g^2(a) a^{6\epsilon}\right),\label{RGE}
\end{equation}
where 
\begin{eqnarray}
A(p) &=& 2\pi D^2 p(p-1)\\
B(p,\epsilon) &=& p(p-1)\left((n-2)\frac{8\pi^2 D^2}{3\epsilon}(1+\epsilon K) + 
\left(\frac{8\pi^2 D^4}{3\epsilon} - \frac{\pi^2}{6}\right)\right) \nn\\
  && \,\,+ p(p-1)(p-2) \left(\frac{\pi^2 D^2}{3\epsilon}(8+\epsilon\alpha) + (p-3)
\frac{4\pi^2 D^4}{3\epsilon}\right).
\end{eqnarray}

There is also a renormalization of the coupling constant $g$.  Calculations were originally presented in \cite{DPP}; we shall not review them here. For a given cutoff $a$, $g$ renormalizes as (tilded operators will represent renormalized quantities)
\begin{equation}
\tilde{g} =  a^{3\epsilon} (g+4\pi g^2 \frac{a^{3\epsilon}}{3\epsilon}),
\end{equation}
with the cut-off dependent factor introduced to return to the cut-off scale $a=1$.  We now invert the renormalization equation up to the second order in $g$:
\begin{eqnarray}
g=a^{-3\epsilon}(\tilde{g} - \frac{4\pi}{3\epsilon}\tilde{g}^2) \\
Z=\tilde{Z}\left(1-\frac{4D^2}{3\epsilon}p(p-1) \tilde{g}\right).
\end{eqnarray}
Replacing bare quantities by renormalized ones in (\ref{RGE}), and using the fact that $D=\frac{1}{2}+{\cal O}(\epsilon^2)$, one gets (we let $g\rightarrow \frac{g}{4\pi}$)
\begin{equation}
\frac{d\tilde{Z}(\xi)}{d\xi} = \tilde{Z}(\xi) p(p-1) \left(\frac{1}{8} g(\xi) + \left((n-2)\frac{1}{48} K - \frac{1}{96} +(p-2)\frac{1}{192}\alpha\right)g^2(\xi)\right),\label{RGS}
\end{equation}
where, we recall, $K=6\log 2$ and $\alpha = 33-\frac{29\sqrt{3}\pi}{3}$.

We can now easily solve the RG equation (\ref{RGS}).  It can be rewritten in the form (dropping the tildes),
\begin{eqnarray}
\frac{dZ(\xi)}{d\xi} &=& \gamma(\xi) Z(\xi) \\
\gamma(\xi) &=& p(p-1)\left(\frac{1}{8} g(\xi) + \left((n-2)\frac{1}{48} K - \frac{1}{96} +(p-2)\frac{1}{192}\alpha\right)g^2(\xi)\right).
\end{eqnarray}

To compute the correlation functions, it will be useful to assume the RG evolution to go from the lattice cut-off ($\sim 1$) to the scale $R$ (we write $\xi_R\equiv \log R$).  To do so, we need the fixed point value of $g$, which we will note $g_*$.  It is know to be of the form \cite{Lu,DPP}:
\begin{equation}
g_* = \frac{3}{2}\epsilon + \frac{9}{4}\epsilon^2 + {\cal O}(\epsilon^3).
\end{equation}
Taking the limit on the number of replicas ($n=0$) and using the explicit form of $g_*$, one obtains the fixed point value of $\gamma$, noted $\gamma_*$
\begin{equation}
\gamma_* = \frac{9}{32}p(p-1)\left(\frac{2}{3}\epsilon + \left(\frac{11}{12}-\frac{2 K}{3}+\frac{\alpha}{24}(p-2)\right)\epsilon^2\right) + {\cal O}(\epsilon^3).
\end{equation}

We are now able to compute the correlation functions.  Using scaling laws, we get
\begin{eqnarray}
\cor &=& \lim_{n\rightarrow 0} \frac{(n-p)!}{n!\,p!}  \langle \sum_{a_1\ne \cdots \ne a_p}^n \sigma_{a_1}(0)\cdots \sigma_{a_p}(0) 
\sum_{b_1\ne \cdots \ne b_p}^n \sigma_{b_1}(R)\cdots\sigma_{b_p}(R) \rangle\nn\\
&\sim& \lim_{n\rightarrow 0} \frac{(n-p)!}{n!}\sum_{a_1\ne a_2 \cdots \ne a_p} (Z(\xi_R))^2 \frac{1}{R^{2p\Delta_{\sigma}}}\nn \\
&\sim& \frac{(Z(\xi_R))^2}{R^{2p\Delta_{\sigma}}}.
\end{eqnarray}
The final result is obtained by using the fixed point value $Z(\xi_R) \sim e^{\gamma_* \xi_R} = R^{\gamma_*}$.  One thus gets
\begin{equation}
\cor \sim \frac{1}{R^{2\Delta_{\sigma^p}'}}.
\end{equation}
with
\begin{equation}
\Delta_{\sigma^p}' = p\Delta_{\sigma} - \gamma_*.
\end{equation}
The deviation from the pure model is thus given by $\gamma_*$. Having show this quantity to be of the form $A\epsilon + B\epsilon^2 + {\cal O}(\epsilon^3)$, we can now look at the domain of validity of the $\epsilon$-expansion.  Evidently, it becomes absurd if $|A\epsilon + B\epsilon^2|\sim |p\Delta_{\sigma^p}|$.  For the 3-states Potts model, this happens for $p\ge 5$.  This explains why this method cannot predict disorder-averaged moment for such $p$'s.  In contrast the expansion makes a good approximation for $p\le 4$.  

To conclude, let us derive another physically interesting quantity, which is the derivative of $\Delta_{\sigma^p}'$ with respect to $p$, evaluated at $p=0$ (it is $\alpha_0$ (not to be confused with the Coulomb gas parameter) in Ludwig's notation).  It describes the asymptotic decay of the spin-spin correlation function ($\langle \sigma(0)\sigma(R)\rangle \sim \frac{1}{R^{2\alpha_0}}$).  It is straightforwardly shown to be
\begin{equation}
\alpha_0\equiv \left(\frac{\partial\Delta_{\sigma_p}'}{\partial p}\right)_{p=0} = \Delta_\sigma + \frac{9}{32}\left(\frac{2}{3}\epsilon + \left(\frac{11}{12} - \frac{2 K}{3} - \frac{\alpha}{12}\right)\epsilon^2\right)+{\cal O}(\epsilon^3).
\end{equation}
This quantity is probably the easiest to measure in numerical simulations.

%-------------------------------------------------------------------
\stars
%--------------------------------------------------------------------
I would like to thank Vl. S. Dotsenko and P. Simon for their suggestions and their help to get used to the integral calculations.  This research was supported in part by the NSERC Canada Scholarship Program and by the Celanese Foundation.

%--------------------------------------------------------------------

\end{document}